\documentclass[9pt,conference]{IEEEtran}
\usepackage{amssymb,amsthm,amsmath,array}
\usepackage{graphicx}
\usepackage[caption=false,font=footnotesize]{subfig}
\usepackage{xspace}
\usepackage[sort&compress, numbers]{natbib}
\usepackage{stmaryrd}
\usepackage{xcolor}
\usepackage{mathtools}
\usepackage{float}
\usepackage{textcomp}
\usepackage{algorithm}
\usepackage{algorithmic}

\begin{document}

\title{A Practical Exercise in Adapting SIFT Using FHE Primitives}
\author{
    \IEEEauthorblockN{
        Ishwar B Balappanawar\IEEEauthorrefmark{1}, 
        Bhargav Srinivas Kommireddy\IEEEauthorrefmark{1}
    }
    \IEEEauthorblockA{
        IIIT Hyderabad \\
        \texttt{\{ishwar.balappanawar, bhargav.srinivas\}@students.iiit.ac.in} \\
    }
    \textit{\IEEEauthorrefmark{1} These authors contributed equally to this work.}
}

\maketitle
\begin{abstract}
An exercise in implementing Scale Invariant Feature Transform using CKKS Fully Homomorphic encryption quickly reveals some glaring limitations in the current FHE paradigm. These limitations include the lack of a standard comparison operator and certain operations that depend on it (like array max, histogram binning etc). We also observe that the existing solutions are either too low level or do not have proper abstractions to implement algorithms like SIFT. 

In this work, we demonstrate:  
\begin{itemize}
    \item Methods of adapting regular code to the FHE setting.  
    \item Alternate implementations of standard algorithms (like array max, histogram binning, etc.) to reduce the multiplicative depth.  
    \item A novel method of using deferred computations to avoid performing expensive operations such as comparisons in the encrypted domain.  
\end{itemize}

Through this exercise, we hope this work acts as a practical guide on how one can adapt algorithms to FHE.
\end{abstract}

\section{Introduction}
How good is the current Fully Homomorphic Encryption (FHE) paradigm at performing common everyday algorithms? To explore this question, we started working on adapting a common image processing algorithm, Scale Invariant Feature Transform (SIFT) \cite{sift}. Fully Homomorphic Encryption (FHE) schemes have been increasingly utilized for secure computations \cite{survey_FHE, chhabra_survey}, but adapting algorithms like SIFT comes with unique challenges. While frameworks like Microsoft SEAL, HElib, and TenSEAL \cite{seal, helib, tenseal} provide strong primitives, their abstractions often fall short for complex algorithms.

In the context of FHE, this task involves two main parties: - a) the client who has the raw image and is the entity that encrypts the image. b) The server which executes the algorithm that needs to take an encrypted image, i.e., a 2-dimensional array of encrypted pixel values and produce the encrypted version of the output. The goal is to ensure accurate execution of the algorithm without revealing the image or intermediate results (like SIFT keypoints) to the processing entity. 

There is existing work on implementing SIFT using homomorphic encryption schemes, but most of the approaches involve revealing some information either about the image or the SIFT points.\cite{reveal_information} \cite{another_secure_sift} We wanted to avoid revealing any kind of information to the processing entity. We also avoid utilizing the internals of the FHE scheme and only use the primitives provided by libraries. This is a deliberate choice made to assess the quality of abstractions that are available. First we will discuss the SIFT algorithm itself and isolate the parts of it that are challenging/interesing to adapt to the FHE domain. In the subsequent sections we will talk about the deficiencies of the state of the art and propose some new ways to deal with the challenges using deferred computations.

\section{The SIFT Algorithm}
The SIFT algorithm involves the following steps\cite{sift_explanation}:
\begin{itemize}
    \item Scale-space Extrema Detection  
    \item Keypoint Localization  
    \item Orientation Assignment  
    \item Keypoint Descriptor Generation  
\end{itemize}

Our goal was to implement all these computational tasks using the primitives provided to us for a FHE scheme. Some of the tasks involve computing gradients and hessians which warrant the use of floating point numbers. For this reason we chose to use the CKKS encoding scheme\cite{CKKS}. We used the tenseal python library, as it is a well supported wrapper around the popular Microsoft SEAL library. The primitives that we utilize are addition and multiplication. So we implemented all the necessary computational tasks using addition and multiplication. There is the further constraint of multiplicative depth, which prevents us from performing a long chain of multiplications on an encrypted number. This is one of the most significant hurdles with adapting any algorithm to the FHE setting. 

Some of the tasks, such as convolution and matrix multiplication, were trivial to implement using addition and multiplication.\cite{inspiration} However there were multiple tasks that were far from trivial, we will discuss these in the next section.

\section{The Hard Tasks} \label{sec:hard_tasks}
\subsection{Comparison}

The computation of scale-space extrema in a Fully Homomorphic Encryption (FHE) setting requires comparing encrypted values, a challenging task due to the complexity of implementing comparison functions. Integer domain comparisons can leverage Fermat's Little Theorem for exact results, while approximate methods exist for floating-point numbers \cite{cheon_comparison}. However, both approaches are resource-intensive, requiring significant multiplication depth, which is costly in FHE.

A practical solution involves an interactive approach where the server sends the two numbers to the client for comparison, and the client returns an encrypted boolean result with minimal noise \cite{juvekar2018copra}. While this method introduces interactivity and potential algorithm exposure (as the client sees the values being compared), the risk can be mitigated by sending spurious comparison requests alongside the real ones. This approach balances practicality with security, leveraging modern communication speeds to maintain efficiency.

\subsection{Division}
Integer division is surprisingly hard to perform using FHE primitives, it often requires the use comparison function which itself is expensive. Floating point division is a bit simpler; approximations such as Taylor or Chebyshev series are commonly used in FHE computations \cite{bootland_polynomial}. To get a good approximation we still need to use excessive multiplicative depth.

\[
\frac{1}{x} \approx \frac{1}{a} - \frac{(x - a)}{a^2} + \frac{(x - a)^2}{a^3} - \frac{(x - a)^3}{a^4} + \cdots
\]

To avoid this, we store the numerator and denominator of the division separately and modify the subsequent uses of the algorithm to use the numerator and denominator separately. For example, if \( c = \frac{a}{b} \) and later we use it for, say, comparison, \( c < d \), we modify this comparison to \( a < d \cdot b \). This allows us to avoid calculating the inverse of \( b \), which would have been expensive.

\subsection{Other Functions Requiring Approximation}
There are steps in SIFT which require the computation of the magnitude of vectors. This involves computing the square root of a number. Again there are polynomial approximations of this using Chebyshev polynomials (used by openFHE). Any sort of polynomial approximation would require a lot of multiplicative depth to get a good approximate value, so we do the same that we did in comparison and delegate this calculation to the client.

\subsection{Conditional Blocks}
The result of a comparison is often used to decide which code block to execute. In algorithms implemented with FHE primitives, we do not have a luxury to choose. If the boolean value of an if else condition is encrypted then we need to execute both the if-block and the else-block and mask the effects of the block by multiplying with the boolean value.

This is branchless coding and there has already been a lot of research done in converting regular algorithms to their branchless version \cite{brodal2001cache, gilad2016cryptonets}. A big reason for this is that to take advantage of the min-maxed parallelism of GPUs, one needs to reduce the number of branches. We believe this field is worth consulting before one starts implementing their algorithms. 

Some optimizations that help in regular algorithms can become unnecessary when using FHE, because we execute both if and else block anyway. So early stopping strategies like breaking in loops is pointless. This allowed us to skip a very large chunk of code in SIFT. One should greedily look for such blocks to eliminate while adapting their algorithms.

\subsection{Histograms and Binning}
In one of the steps we need to calculate the magnitude and angle of gradients in a neighborhood. A histogram of these angles need to be calculated weighted by their magnitudes. An efficient representation of the histogram bins is critical. This kind of optimization has been explored in privacy-preserving image processing \cite{privacy_preserving_sift}. Typically, we would implement this by creating an array with 360 / num\_bins entries. For a given angle, one would calculate the index of the array that needs to be incremented. In the FHE setting we cannot choose which element of the array to increment, we need to update every element but multiplied with a mask so that only the correct element is incremented. The mask is generated by comparing the candidate element with the limits of each of the bins. \( a_i < \theta < a_{i+1} \). for all $i$. This means that indices in FHE space are one-hot vectors. 

Before we start calculating the histogram, we need to calculate the angles of the gradient according to the following equation.

\[
\theta = \arctan\left(\frac{dy}{dx}\right)
\]

We have the values for dx and dy but using approximations to calculate inverse of dx and arctan is just too expensive, so we modify the condition we use to calculate the mask.

\[
\tan(a_i) \cdot dx < dy < \tan(b_i) \cdot dx
\]

Now we have the bins in terms of multiplications and additions. There is still a tan to calculate but that is over an unencrypted value, so can do it regularly. Would a simple compiler pass of the regular histogram binning code be able to do an optimization like this? Probably not. Is it impossible to come up with an automated way of doing this? Probably not. The conversion here needed knowledge of the properties of arctan and tan, but nothing more. This opens up new directions to look into in terms of automatic conversion of regular algorithms to code that works well using FHE primitives.

\subsection{Finding Max in an Array}
In the step of orientation calculation, one needs to find the maximum element in the histogram. The usual method of doing this would be maintaining a running maximum of the elements and comparing that running maximum with subsequent elements. Update the running maximum according to the boolean result of the comparison.

\begin{algorithm}
\caption{Find Maximum in Array}
\begin{algorithmic}[1]
\STATE $max \gets 0$
\FOR{each element in array}
    \STATE $b \gets (element > max)$
    \STATE $max \gets b \cdot element + (1 - b) \cdot max$
\ENDFOR
\end{algorithmic}
\end{algorithm}

If we analyze the multiplicative depth of this algorithm, it would be O(N), where N is size of array, ignoring the requirement for comparison.

Since length of arrays can be large, we instead use a different way of calculating max, where we compare adjacent elements of the array and remove the smaller values. This would cut down the array size in half. We perform this step till only one element is left. The multiplicative depth of this algorithm ignoring comparison is O(logN).

\begin{algorithm}
\caption{Max Function}
\begin{algorithmic}[1]
\STATE \textbf{function} \textsc{Max}($a, b$)
\STATE \hspace{1em} $cond \gets (a > b)$
\STATE \hspace{1em} \textbf{return} $cond \cdot a + (1 - cond) \cdot b$
\end{algorithmic}
\end{algorithm}

\begin{algorithm}
\caption{Vector Maximum (\textsc{VecMax})}
\begin{algorithmic}[1]
\STATE \textbf{function} \textsc{VecMax}($ls$)
\STATE \hspace{1em} $l \gets \text{length of } ls$
\STATE \hspace{1em} \textbf{if} $l = 1$ \textbf{then}
\STATE \hspace{2em} \textbf{return} $ls[0]$
\STATE \hspace{1em} \textbf{else if} $l = 2$ \textbf{then}
\STATE \hspace{2em} \textbf{return} \textsc{Max}($ls[0], ls[1]$)
\STATE \hspace{1em} \textbf{else}
\STATE \hspace{2em} \textbf{return} \textsc{Max}(\textsc{VecMax}($ls[:l//2]$), \textsc{VecMax}($ls[l//2:]$))
\end{algorithmic}
\end{algorithm}

\section{Deferred Computation} \label{sec:deferred}

In order to delegate expensive operations, such as comparison and square root, to the client in a non-interactive manner, we propose a novel method of \textbf{deferring computation}. The deferred computation technique is inspired by concepts from dynamic computation graphs and secure function evaluation, as used in frameworks like TensorFlow for privacy-preserving tasks \cite{tensorflow_computation_graph}. In this method, the server sends a single response to the client. The client can then perform the requisite computation to extract the result from the response.  

\subsection{How It Works}  

The server assumes it does not know the boolean value and treats it as a variable. As operations are performed on this boolean variable, a function is constructed with the variable as a parameter. The server sends the following to the client:  
\begin{itemize}  
    \item The comparison required to compute the boolean value.  
    \item A function that uses this boolean value as a parameter.  
\end{itemize}  

The client performs the necessary comparison and substitutes the resulting values into the function to compute the final result.  

\subsection{Construction of the Function}  

To construct this function, a \textbf{dynamic computation graph} is created, which tracks how the comparison results are manipulated. This graph contains comparison nodes and represents the structure of the computation. Post-processing may be applied to simplify the graph and produce a more compact function.  

\subsection{Example}  

Consider the following expression:  
\[
a = (x > y) \cdot c + (z > w) \cdot d + (y \geq x) \cdot e
\]  
The server will return to the client:  
\[
f(c_1, c_2) = c_1 \cdot (c - e) + c_2 \cdot d + e
\]  
where:  
\[
c_1 = (x > y), \quad c_2 = (z > w)
\]  

\subsection{Downside of the Approach}  

One potential downside of this approach is that, in the worst case, the client could extract the exact algorithm used by the server. For example, if the entire algorithm depends on a single initial comparison, the computation graph will store all the subsequent computations. Thus, the complexity of the function depends heavily on the algorithm itself. However, in most cases, the resulting function is likely to be simpler.  

\section{Final Thoughts}

Through this exercise we see that there is an absence of a framework that can be used to implement general algorithms \cite{gilad2016cryptonets}. While execution time might be a problem, applications such as CNNs and Neural Networks are trivial to implement given the primitives.The vast majority of libraries observed either provide only basic primitives or attempt automatic code translation with limited success \cite{survey_FHE} \cite{survey}. Frameworks like SEAL, HElib, and OpenFHE \cite{seal, helib, openfhe} are excellent for secure computations but lack intermediate abstractions for algorithm adaptation.  The former is not friendly to beginners who are trying to use the technology but don't want to know the details of the field while the latter is inconsistent in how it works. 

There needs to be an intermediate framework which abstracts what can trivially be abstracted away but still make it explicit where the user needs to be careful, so that the user knows the tradeoffs of using a particular feature. If we want practicality, then we need to embrace the limitations of the current FHE scene.

\bibliographystyle{IEEEtran}
\bibliography{references}

\end{document}